# Space-Time Quantization, Elementary Particles and Dark Matter

A. Meessen

*Institute of Physics, Université Catholique de Louvain, Louvain-la-Neuve, B-1348*

*Email: auguste@meessen.net*

**Abstract.** Relativity and quantum mechanics are generalized by considering a finite limit for the smallest measurable distance. The value *a* of this *quantum of length* is unknown, but it is a universal constant, like *c* and *h*. It depends on the total energy content of our universe (hc/2a) and physical laws are modified when it is finite. The eigenvalues of (x, y, z, ct) coordinates are integer or half-integer multiples of *a*. This yields *four new quantum numbers,* specifying "particle states" in terms of phase differences at the smallest possible scale. They account for all known elementary particles and predict the existence of neutral ones that could constitute dark matter particles. This theory is thus experimentally testable.

## Introduction

The familiar concept of a "space-time continuum" implies that it should be possible to measure always smaller and smaller distances without any finite limit. Heisenberg, who insisted on expressing quantum mechanical laws in terms of measurable observables, questioned already the validity of this postulate [1]. We should thus treat the ultimate limit *a* for the smallest measurable length as a yet unknown quantity. Actually, we learned already from the development of relativity and quantum mechanics that Nature can impose restrictions on our measurements because of two universal constants: the velocity *c* and the quantum of action *h.* Could Nature impose a third restriction, resulting from the existence of a universally constant *quantum of length a* and a universally constant *quantum of time a/c?*

To answer this question, we replaced differential field equations by corresponding finite difference equations. This led to *a generalization of Einstein's energy-momentum relation*, implying fundamental changes as soon a ≠ 0, even when the actual value of *a* is extremely small [2]. Superluminal velocities are only excluded, for instance, when a = 0. Nevertheless, there are no logical in consistencies, since Space-Time Quantization (STQ) preserves relati-vistic invariance, as well as causality [3]. It provides also a simple and natural way to distinguish elementary particles from one another [4] and predicts the existence of new types of elementary particles. The validity of this theory can thus be *tested,* by checking its predictions at great accelerators, like the LHC, and by means of satellite-borne cosmic ray experiments.

Since any generalization of existing theories allows for degrees of freedom, we had to justify our basic choices with respect to other propositions [4]. Although the idea of a minimal length is now more easily acceptable, it is still used in various ways [5]. This field of research is reactivated today, since it concerns the foundations of physics and since we become more and more aware of many unsolved problems. We will briefly review the basic ideas of STQ and apply them in an improved way to elementary particle physics and general relativity. Justification and extension of the standard model of elementary particle physics, identification of dark matter particles and removal of the Big Bang singularity belong, indeed, to the most prominent problems of present day physics.

## Generalized space-time coordinates

We postulate that the value $a$ of the smallest measurable distane is *a universal constant for any, arbitrarily chosen direction.* In principle, we could thus perform ideally precise distance measurements by means of successive juxtapositions of this quantum of length. If such measurements were performed in an inertial reference frame along a given x-axis by starting at its origin, the spectrum of possible eigenvalues would thus be x = 0, ±a, ±2a, … This is the *normal* spectrum of x, including the chosen origin. However, when we assume only that *x* is a possible result, -*x* is also one, since the orientation of the x-axis is arbitrary. The measurable distance 2*x* has thus to be an integer multiple of *a.* It follows that there exists also an *inserted* spectrum, where x = ±a/2, ±3a/2, …

This result is analogous to the quantization of the generalized angular momentum component $J_z$ along any given z-axis in terms of integer or half-integer multiples of $\hbar$. However, the three components of an angular momentum vector are defined in terms of position and momentum observables and are thus not simultaneously measurable with absolute precision. This is different for (x, y, z, ct) space-time coordinates, since they are determined by *independent measurements.* The usual postulate that space and time have 3+1 dimensions remains valid at any scale, and the axes for different space-time measurements can be chosen in a completely arbitrary way. For any Cartesian inertial reference frame, the eigenvalues of the four space-time coordinates are thus always integer or half-integer multiples of *a.*

*Only measured distances are quantized.* They are parallel to the chosen reference axes, while all other lengths can be calculated in terms of measured ones. They don't have to be integer or half integer multiples of *a,* since this is not an "atom of length", requiring commensurability. Moreover, STQ preserves the *isotropy and homogeneity of space and time,* since the origin and orientation of the reference axes can be chosen in a completely arbitrary way.

**The generalized energy-momentum relation**

Let us consider a point-like material particle or the centre of mass of any system of rest mass $m_o$. When it is freely moving along a given direction, we can choose this direction as the x-axis of an inertial reference frame. All possible "states of motion" of this particle are then specified in relativistic mechanics by means of two classically defined observables: the momentum *p* and the energy *E,* which are subjected to *Einstein's relation*

$$(E/c)^2 = p^2 + (m_o c)^2 \qquad (1)$$

Quantum mechanics defines *new* observables *p* and *E* in terms of the wave function

$$\psi = A\, e^{i(kx-\omega t)} \qquad \text{where} \qquad p = \hbar k \quad \text{and} \quad E = \hbar \omega \qquad (2)$$

The values of *p* and *E* are said to be sharply defined for such a harmonic function. Linear superpositions of functions like (2) yield wave-packets, subjected to Heisenberg's uncertainty relations, but for any wave function $\psi$, we can also define "local values" of p and E at a given point and a given instant. They are those values that would yield there the same variations of $\psi$ *at the smallest possible scale* than the harmonic function (2). In the usual continuum theory, this is equivalent to the statement that the partial derivatives

$$\partial_x \psi = ik\psi \qquad \text{and} \qquad \partial_t \psi = -i\omega\psi \qquad (3)$$

Asserting that (1) remains valid for these local values of E and p, we get *the Gordon-Klein equation.* It involves second order partial derivatives $\partial_x^2$ and $\partial_{ct}^2$, but when $a \neq 0$,

$$\partial_x^2 \psi \;\to\; D_x^2 \psi = \frac{\psi(x+a,t) + \psi(x-a,t) - 2\psi(x,t)}{a^2} = \frac{-\sin^2(ka/2)}{(a/2)^2}\psi \qquad (4)$$

at any arbirarily chosen point x for the expression (2). A similar definition of the operator $D^2_{ct}$ allows us to treat E/c in the same way. When $a \neq 0$, the Gordon-Klein equation becomes thus a finite difference equation that is equivalent to

$$\sin^2(\pi aE/hc) = \sin^2(\pi ap/h) + \sin^2(\pi am_oc/h) \qquad (5)$$

This is a new physical law, generalizing Einstein's relation (1) by taking into account the fact that *local values of p and E/c cannot be defined at a smaller scale than a.* The last term of (5) preserves the usual definition of the rest energy: $E = m_oc^2$ when $p = 0$. The generalized energy-momentum relation reduces to (1) when $a = 0$, but also when $a \neq 0$, as long as the arguments of the sine functions are very small compared to $\pi/2$. The "continuum approximation" requires thus only that E, pc and $m_oc^2$ be very small compared to hc/2a.

Einstein's relation (1) corresponds to hyperbolas, while the more general relation (5) is periodic for p and E/c. When we consider only positive values of E and pc, limitted by $E_u$ = hc/2a, we get curves like those of figure 1. Different "states of motion" are thus defined by values of $\omega/c$ and k between $-\pi/2$ and $+\pi/2$. This corresponds to the first Brillouin zone.

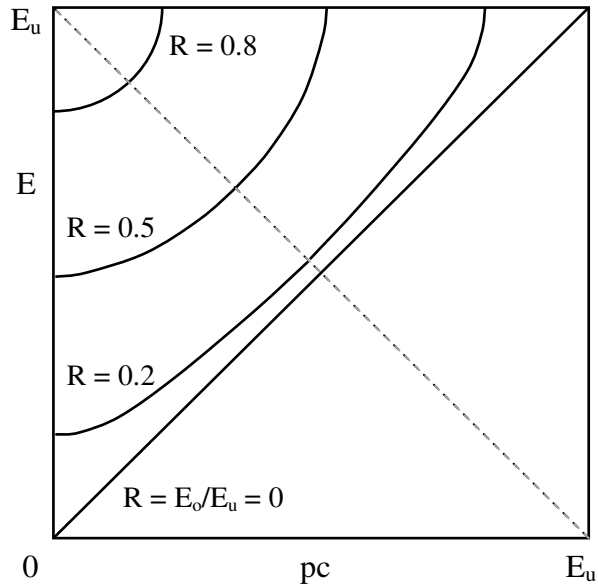

FIG. 1: **Generalization of Einstein's relation.** The rest energy $E_o = m_oc^2$ and $E_u$ = hc/2a.

The curves shrink when the rest energy $E_o$ tends towards $E_u$. The limit is a single state, where $p = 0$ and $E = E_u$. This corresponds to rest in the quantum-mechanical sense and allows us to assert that *$E_u$ is the total convertible energy content of our universe.* Indeed, if a material body had a rest energy $E_o = E_u$, there would be no energy left that could appear in the form of kinetic energy. Customary physical laws have then to be modified for energies that are approaching the highest possible value, when $E_u$ is finite. Since $|p| = h/\lambda \leq h/2a$, the smallest possible scale for interference measurements would be $\lambda/2 = a$.

The limitation of the energy-momentum space for possible states of motion is irrelevant under normal circumstances, but it solves a basic problem concerning *the definition of inertial frames.* They are those frames where the principle of inertia is valid and where material bodies can only be accelerated when they are subjected to *real forces,* resulting from physical interactions with something else. This led Ernst Mach to ask: *"how does it come about that inertial systems are physically distinguished above all other coordinate systems?"* Since the principle of inertia should be valid in the whole universe, he suggested that inertial frames have to be un-accelerated with respect to "fixed stars". He actually meant the whole universe. Einstein recognized that the theory of relativity did not justify such a relation [6], but STQ solves this riddle in a self-consistent way.

**Superluminal velocities and relativistic invariance**

We describe *motions* of the center of mass of any material system by considering the evolution of its quantum-mechanical probability distribution. The classical velocity v is then the group velocity of the associated wave function. It is the velocity of the point where interference of superposed harmonic functions like (2) leads to a maximum. At this point, the components of the $\psi$ function have identical phase factors, so that $x = x_o + vt$, where the velocity $v = d\omega/dk = dE/dp$. This remains valid when $a \neq 0$, since the *calculated* average position does not have to coincide with some lattice-point.

The slope of the curves in figure 1 defines the ratio v/c. Photons are thus always moving in vacuum at the velocity v = c, while free material particles can move at any velocity. *Superluminal velocities are not forbidden anymore when $a \neq 0$.* The "light barrier" for material particles corresponds to the second diagonal in figure 1. Since superluminal velocities require energies $E > E_u/2$, they are irrelevant for practical purposes, but they insure the internal consistency of physical theories. Indeed, Einstein noted that a $\psi$-function can be defined for two particles in a specially prepared (entangled) state, so that a measurement performed on one particle will instantaneously provide information about the other particle, however great their separation may be [7]. This would be in conflict with special relativity, but follows from the fact that a $\psi$ function defines the *knowledge* we have about a given system. It is modified everywhere in space at the very instant where a new measurement is performed.

Even photons can move at superluminal velocities, when they interact with a medium. The group velocity of light can even be more than 100.000 times larger than c [8]. Because of the lack of perfect consistency between relativity and quantum mechanics, Einstein suggested that the quantum-mechanical description of reality could be incomplete. STQ provides a generalization of quantum mechanics, but *the special theory of relativity is also modified* for classical (quantum-mechanical average) motions.

This appears in particular when we consider the *usual Lorentz transformation.* It applied to possible values of space-time coordinates in different inertial frames, but insured only the constancy of c. It was implicitly assumed that a = 0, while the value of h was irrelevant. Actually, we have to require the invariance of the new physical law (5), including c, h and a. It is sufficient to replace the values of E and p appearing in (1) by the corresponding sine-functions in (5), to get a *generalized Lorentz transformation* in energy-momentum representation [3]. This transformation law is deterministic, since the eigenvalues of E and p have a continuous spectrum of possible values in a quasi-infinite universe. The resulting velocity addition law *allows for superluminal velocities, without contradicting the principle of causality.* It applies to calculated average positions, while ideally exact measurements of space-time coordinates yield quantized eigenvalues with the same spectrum for all inertial frames.

Newton's law of motion results from the definition of the average acceleration dv/dt = $(d^2E/dp^2).(dp/dt)$ = F/m, where F = dp/dt defines the applied force and m the *inertial mass* of the material system. To describe motions along any direction with respect to three arbitrarily chosen reference axes, we need three momentum components. The energy-momentum relations (1) and (5) are then generalized by introducing a sum over $p_j$, where the index j = x, y, z. The inertial mass becomes a tensor, with velocity dependent components.

At this stage, a scientific meeting [9] provided the opportunity to submit all these arguments to *Werner Heisenberg.* In his answer [10], he formulated no objections and advised even to search for experimental confirmations. He didn't say how, but we were already trying to find out if a ≠ 0 could account for the emerging spectroscopy of elementary particle physics, as h ≠ 0 did for atomic physics. Since Dirac discovered some internal degrees of freedom of elementary particles, when he replaced the second order differential Gordon-Klein equation by an equivalent set of first order differential equations, we tried to apply the same procedure to finite-difference equations. This led to the appearance of new degrees of freedom and made us aware of the existence of "generalized space-time coordinates". They allowed for a simpler and more direct approach.

**Four new quantum numbers**

The definition (4) of second order finite derivatives applies to any space-time lattice of lattice constant *a.* It could be an inserted lattice, as well as the normal lattice. There has thus to exist a Ψ function that is defined on all these lattices with a smoothly varying probability distribution $|\Psi|^2$. It could be bell-shaped, for instance, but the Ψ function can oscillate with *different phase factors* on intercalated lattices with respect to the normal lattice.

To justify this assertion for a given x-axis, we consider the function $\varphi(x) = e^{iKx}$, where x can be an integer or half-integer multiple of a. The *generalized momentum* of the particle is $\hbar K$, where $K = k + u(2\pi/a)$ in the extended zone scheme. The usual momentum $p = \hbar k$, where k belongs to the first Brillouin zone ($-\pi/a \leq k \leq +\pi/a$). This value of k defines possible "states of motion", but we can consider various Brillouin zones, specified by $u = 0, \pm 1, \pm 2$ ... Using the resulting value of K, we get always $\varphi(x) = e^{ikx}$ for the normal lattice, since x is then an integer multiple of a, but for the inserted lattice, $\varphi(x) = e^{iu\pi}e^{ikx}$. The quantum number u accounts thus for a degree of freedom that did not appear as long as we believed that space and time had to be continuous. The concept of an inserted lattice allows us to define the new quantum number u for any wave function $\varphi(x)$. We have to apply a translation operator at the smallest possible scale with respect to any given point x, but it is not sufficient to consider only $\varphi(x+s)$, where $s = \pm a/2$, since there are possible phase differences. Thus,

$$T_x^{\pm} \varphi(x) = U_x \varphi(x \pm a/2) \qquad \text{where} \qquad U_x = \exp(iu_x\pi) = \pm 1 \qquad (6)$$

We add an index x to U and u, since the same procedure applies to all (x, y, z, ct) coordinates. This yields *four new quantum numbers* ($u_x$, $u_y$, $u_z$, $u_{ct}$) that can only have integer values. They define "particle states", even for true elementary particles. Since they can't have some internal structure, distinguishing them from one another, they are single points. Nevertheless, they are distinguishable by means of different variations of their wave functions in space and time. They can be surrounded by a cloud of virtual particles, determining the rest mass of the observable entity, but the possible types of virtual particles depend also on the core particle.

In classical and relativistic mechanics, we imagined perfectly localizable point-particles, moving along well-defined trajectories. Quantum mechanics introduced the concept of propagating wave functions, but the type of particles had still to be specified in a phenomenological way. It was an enourmous achievement to recognize that *elementary particles are characte-* *rized by a set of quantized observables,* but the reason for their existence remained obscure. This happened also for atomic quantum numbers at the beginning of the 20th century. Their physical origin could only be explained by constructing quantum mechanics, where the quantum of action h ≠ 0. String theories tried to justify the existence of new degrees of freedom by assuming additional dimensions for space and time, but this is not necessary when a ≠ 0.

Different types of elementary particles are now characterized by their ($u_x$, $u_y$, $u_z$, $u_{ct}$) quantum numbers, specifying how the associated Ψ functions vary in space and time at the scale of a/2. In this sense, we can say that *particle states correspond to different patterns of excitations of space and time*. For a given type of particles, this pattern is always and everywhere the same in the whole universe. It is also the same for any reference frame, since the u-quantum numbers are always specified by the phase factors of wave functions on inserted lattices with respect to the corresponding normal lattice. This expresses again the basic idea of relativistic invariance: although we need reference frames for actual measurements, the resulting physical laws are independent of the chosen reference frame.

Quantum mechanics allowed us to describe possible "states of motion" of a given particle in a much more detailed way than in classical physics, by considering wave functions in space and time. Now, we get even a more detailed description, by adding *modulations at the smallest possible scale.* This is a natural extension of quantum-mechanical principles, where observable properties of particles are defined in terms of possible variations of their wave functions in space and time. The existence of various types of elementary particles called for new degrees of freedom, but it is not necessary to imagine that elementary particles are strings or branes in some enlarged space. Ordinary, but quantized space and time is sufficient.

The u-quantum numbers did not appear in (5), since finite *second order* derivatives like (4) yield identical results for any particular space-time lattice of lattice constant *a.* To generalize Dirac's procedure in a simple way, we introduce intercalated lattices and the operator

$$D_x = \frac{T_x^+ - T_x^-}{a} \qquad \text{so that} \qquad D_x \varphi(x) = U_x \frac{\sin(ka/2)}{ia/2} \varphi(x)$$

Because of (6), this defines not only the local value of the ordinary momentum $p = \hbar k$ along the x-axis, but also the new observable $u_x$. In the continuum approximation, it is sufficient to generalize Dirac's equation by setting $\partial_x \to D_x \to U_x \partial_x$, but the $U_x$ factor is only justi-fied for STQ. When the properties of *two* freely moving elementary particles are specified by the wavefunctions $\varphi_1(x)$ and $\varphi_2(x)$ for a given x-axis, their ensemble is such that

$$T_x^\pm \, \varphi_1(x) \, \varphi_2(x) = U_x^1 U_x^2 \, \varphi_1(x \pm a/2) \, \varphi_2(x \pm a/2)$$

Since $U_x^1 U_x^2 = U_x = \exp(iu_x\pi)$, where $u_x = u_x^1 + u_x^2$, we get *separate addition laws* for u-quantum numbers along any particular space-time axis. To account for gauge theories, we consider a free particle, having a well-defined momentum along the x-axis. Interactions lead to a modification of the phase factor, so that $\varphi(x) = e^{i(kx+f)}$, where $f = f(x)$. In the continuum approximation, we get then $\partial_x \varphi(x) \to U_x i\left(k + U_x^f \partial_x f\right)\varphi(x)$. The product $U_x U_x^f$ implies that we can consider the sum $u_x + u_x^f$ to characterize these interactions. We get thus *separate conservation laws* for different space-time axes.

## Justification and extension of the standard model

To compare the standard model of elementary particle physics with the implications of STQ, we have to consider particles that are characterized by four ($u_x$, $u_y$, $u_z$, $u_{ct}$) quantum numbers. We start with the sub-group, where $u_{ct} = 0$, while $u_x$, $u_y$ and $u_z$ can be equal to 0, 1 or $\bar{1} = -1$. Considering $u_x$, $u_y$ and $u_z$ axes, we can easily identify some of the resulting lattice points with *known fermions of ordinary matter,* as indicated in figure 2.

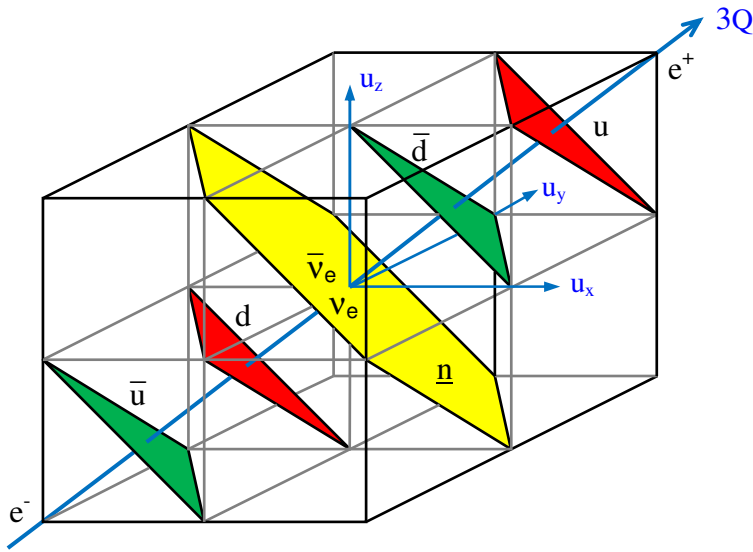

FIG. 2: **Representation of particle states by means of lattice-points.** The $u_x$, $u_y$ and $u_z$ quantum numbers can be equal to 0, ±1, while $u_{ct} = 0$. This accounts for all known leptons and quarks of ordinary matter, but predicts also the existence of neutral quarks $\underline{n}$.

The electric charge q = Qe is quantized in such a way that $u_x + u_y + u_z = 3Q$, so that Q is the average value of the spatial quantum numbers. The positron state (1,1,1) and the electron state $(\bar{1}, \bar{1}, \bar{1})$ correspond to Q = ±1. The up-quark (u) has three possible states: (0,1,1), (1,0,1) and (1,1,0) with Q = 2/3. The down-quark (d) can also exist in three possible versions: $(\bar{1}, 0, 0)$, $(0, \bar{1}, 0)$ and $(0, 0, \bar{1})$ with Q = -1/3. Particle and antiparticle symmetry corresponds to sign reversal of $u_x$, $u_y$ and $u_z$ when $u_{ct} = 0$. Experiments proved that there are only three different "color states". Now, it appears that they result from the three-dimensional nature of space, but the names of the axes are arbitrary. This allows for permutations of $u_x$, $u_y$ and $u_z$.

In figure 2, we represent the triangles for quark states in red and those for anti-quark states in green. Looking along the Q-axis, we get a central point, four triangles and a hexagon. The triangles are superposed in such a way that they yield three (R, G, B) color states for quarks and three $(\bar{R}, \bar{G}, \bar{B})$ anti-color states for antiquarks. The corners of the *hexagon* define new particle states of charge Q = 0. They are of type $(1, \bar{1}, 0)$, with 6 possible permutations of these quantum numbers. They can also be specified in terms of color and anti-color states, like $R\bar{G}$ or $G\bar{R}$ for instance, but what about the (0,0,0) state at the centre of the hexagon?

Previously [4], we considered only two (0,0,0) states for the electron neutrino and electron anti-neutrino, but neutrinos are leptons that belong to the Q-axis, like $e^{\pm}$ states. Now, we add two (0,0,0) states to those of type $(1, \bar{1}, 0)$ to get *a group of 8 states of charge Q = 0*. Since they belong to a plane that is perpendicular to the Q-axis, they are analoguous to quarks and antiquarks. For simplicity, these "neutral quarks" will be called *narks*

Instead of considering only $u_{ct} = 0$, we have also to allow for $u_{ct} = \pm 1$, associated with various sets of ($u_x$, $u_y$, $u_z$) quantum numbers, as in figure 2. We get then *three generations* of spin ½ particles. In figure 3, we represent them in terms of $u_{ct}$ and $Q = (u_x+u_y+u_z)/3$.

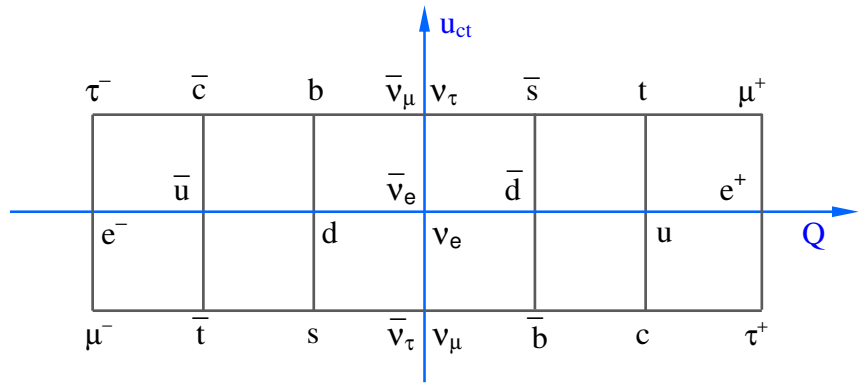

FIG. 3: **The three known generations of quarks and leptons**
are characterized by $u_{ct}$ = 0, ±1 and Q = 0, ±1/3, ±2/3, ±1.

We set $u_{ct} = +1$ for the top quark t, since we can freely choose the orientation of the $u_{ct}$ axis, as this happened also for the Q axis. The sign of all ($u_x$, $u_y$, $u_z$, $u_{ct}$) quantum numbers is opposite for particles and antiparticles. It is now clear that charge conjugation $C = PT$, where P is the parity operator and T the time reversal operator.

## Bosons and simplified SUSY

The value $J_z = m\hbar$ of the generalized angular momentum component determines the variation of the wave function around a given z-axis by means of the factor $e^{im\phi}$. The angle $\phi$ defines the azimuth around the z-axis, while possible values of m are determined by the condition that the *probability* distribution has to be periodic when the angle $\phi$ increases by $2\pi$. This allows for a change of sign of the wave function, so that *m* can be an integer or half-integer number. Since this remains valid for $\Psi$ functions that are defined on the normal and intercalated space-time lattices, *the spin is independent of the u-quantum numbers.*

There are thus spin ½ fermions, characterized by ($u_x$, $u_y$, $u_z$, $u_{ct}$) and spin 1 bosons, characterized by [$u_x$, $u_y$, $u_z$, $u_{ct}$]. Every fermion state corresponds to a boson state and vice-versa. This is known as *supersymmetry* (SUSY), but it is commonly assumed that known fermions and known bosons require the existence of hitherto unknown s-partners. Even if this were true, the masses would be different. It is thus simpler to assume that we know already some super-partners and to represent also boson states as we did for fermions in figure 2 and 3.

Photons are spin 1 particles, characterized by [0,0,0,0], while Z and $W^{\pm}$ bosons are respectively in [0,0,0,0], [1,1,1,0] or [$\bar{1}, \bar{1}, \bar{1}, 0$] states. As in figure 2, these states are represented by points on the Q-axis, so that $W^{\pm}$ bosons correspond to $e^{\pm}$ leptons. The Z particle and the photon correspond to the electron neutrino and its antiparticle, but photons have no rest mass, while Z and $W^{\pm}$ particles have a certain mass. They allow for weak interactions, corresponding to transitions that are *parallel to the Q-axis*. For instance, an up-quark can be converted into a down-quark of the same color by emitting a $W^+$ boson: $(0,1,1) \rightarrow (\bar{1}, 0, 0) + [1,1,1]$. This $W^+$ boson can then be converted into a positron and an electron neutrino: $[1,1,1] \rightarrow (1,1,1) + (0,0,0)$. Electroweak unification acquires thus an intuitive meaning.

The known *8 gluons* correspond to the 8 narks of figure 2 and provide an additional argument for their existence. Gluons are mediators of strong interactions, corresponding to transitions that are *perpendicular to the Q-axis*. A red up-quark can thus become a green up-quark, by emitting a $R\bar{G}$ gluon, or $(0,1,1) \rightarrow (1,0,1) + [\bar{1}, 1, 0]$. Narks are also subjected to strong interactions, since they can be transformed into one another by emitting or absorbing gluons. This applies to all narks and antinarks, color-neutral as well as colored ones.

We can also expect the existence of (X and Y) bosons that are analogous to quarks. They correspond to new types of forces, yielding oblique transitions in figure 2. An energetic quark can thus create a nark and its anti-nark: $(0,1,1) \to (\bar{1},1,0) + [1,0,1]$ and $[1,0,1] \to (0,1,1) + (1,\bar{1},0)$ for instance. Antimatter can be converted into matter and vice-versa. We can also consider various types of Higgs particles, which are spin 0 bosons that contribute to the mass of other particles. Higher values of the u-quantum numbers are also possible.

## Hyperons, dark matter and experimental tests

Nucleons and other baryons are combinations of three quarks in different color states. We can thus represent them by means of dots for occupied states, as in the first diagram of figure 4. Quarks are then bound to one another by emitting and absorbing gluons, which is equivalent to exchanges of neighboring dots. The second graph of figure 4 represents a meson, where a quark and an antiquark are bound together. Emission and absorption of a gluon is then equivalent to shifting the dots in opposite directions. *Narks* can also be bound to one another by exchanging gluons. We get thus various entities, represented in figure 4. The index specifies the number of narks and antinarks that are bound to one another.

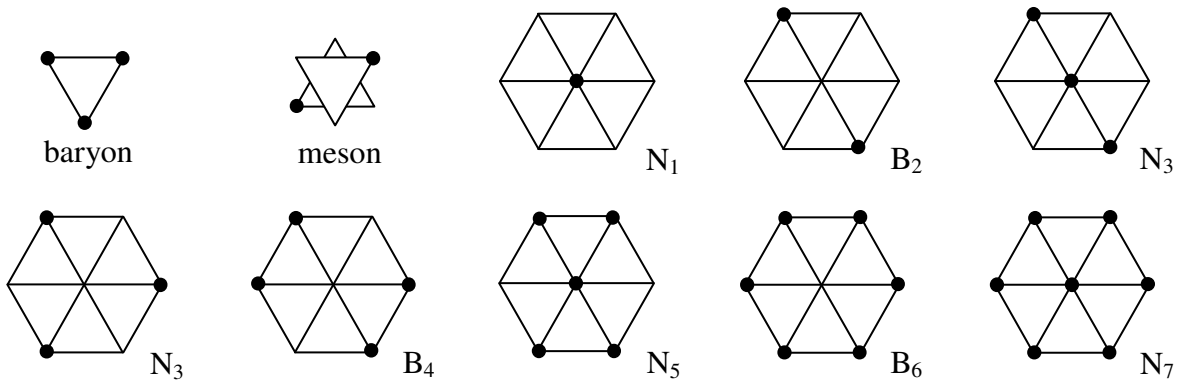

FIG. 4: **Composite particles** are represented by dots for occupied particle states.

$N_1$, $N_3$, $N_5$ and $N_7$ are fermions that are electrically neutral. Since they are analogous to nucleons, we call them "neutralons". $B_2$, $B_4$ and $B_6$ are bosons. They correspond to bound nark-antinark pairs, but are only stable as long as two dots are shifted in opposite directions without leaving the contour of the hexagon. When two opposite dots are shifted towards the centre, the pair can eventually be annihilated. These bosons are metastable.

Narks appeared during the Big Bang, together with quarks and leptons. This led to the formation of *stable nucleons and neutralons of ordinary matter.* Since the binding energy of neutralons increases when the number of interacting pairs is increased, we have to expect *stepwise fusion processes.* They can involve metastable bosons, as for $N_1 + B_2 \rightarrow N_3$ or $N_3 + B_4 \rightarrow N_7$. Neutralons could even be bound to one another like nucleons in nuclear matter, by emitting and absorbing $B_2$ bosons instead of mesons. Atom-like structures and their derivates are impossible for neutral particles, but there could exist various types of neutral entities, consisting of narks and antinarks. Being distributed throughout the whole universe and having some mass, they are good candidates for *dark matter particles.*

Because of spontaneous fusion processes, energy could be released everywhere in our universe, while the mass density is progressively reduced. This could be of cosmic relevance for inflation and the continued accelerated expansion of our Universe. Dark matter could even be a source of energy that is available in interstellar space. However, the essential point is that *it is possible to test the validity of STQ* by verifying whether narks and neutralons do exist or not. Probably, this can be achieved quite soon at the LHC, since high-energy proton-proton and heavy ion collisions could not only create mesons, but also $B_2$, $B_4$ and $B_6$ bosons. Since they are unstable, it should be possible to detect decay products.

Perhaps, one has already observed some phenomena of this type, since the PAMELA satellite experiment on the positron abundance in cosmic radiation between 1.5 and 100 GeV revealed a rising excess at energies above 10 GeV. This cannot be explained by standard models and could be "the first indirect evidence of dark matter particle annihilation" [11]. Moreover, one has observed unexpected multi-muon events in $p\bar{p}$ collisions at the Fermilab collider [12]. Since muons were created outside the beam pipe, having a radius of 15 mm, they had to result from the decay of metastable particles. These mysterious X-particles can decay into lepton-antilepton and quark-antiquark pairs [13], which is possible for $B_4$ and $B_6$ bosons. For the analysis of experimental results, it could thus be useful to know about such possibilities.

## Gravitational effects

Our initial purpose was only to find out whether STQ is possible or not [2], by considering *the ultimate limit* for the smallest measurable distance, without any restriction concerning the type of particles or interactions that might be used. Thus, we treated the value *a* as an unknown quantity, instead of assuming *a priory* that it should be Planck's length L, defined by $L^2 = hG/c^3$ where G is the gravitational constant. Actually, we considered only the logical implications of a finite quantum of length *a,* while its actual value remained unspecified. Inconsistencies would have proven *ad absurdum* that a ≠ 0 is impossible, but this was not the case. On the contrary, it appeared how physical laws have to be generalized when a ≠ 0 and why physics becomes then more logically consitent. This applies even to elementary particles.

Planck's length L was initially considered to show that c and h, combined with G, define universal units for length, time and mass measurements. Later on, it appeared that the metric of general relativity breaks down inside a spherical mass M when its radius is of the order of $r_o = GM/c^2$. Since the Compton wavelength of this mass is $\lambda_o = h/Mc$, we get $L^2 = r_o\lambda_o$, which is independent of the mass M. The length L would thus be *a scale factor for quantum gravity,* as Bohr's radius was for atomic physics. It could determine the size of a fluctuating foamy structure of space and time, for instance, without imposing a lattice-like structure at this scale.

The quantum of length *a* is smaller than Planck's length L, since its value is determined by the total energy content of our universe. Moreover, it has to be considered everywhere, even in those regions where gravitational effects are negligible, while quantum gravity calls for extremely intense gravitational fields. Nevertheless, the quantum of length *a* is relevant for cosmology, since narks and their associations seem to account for dark matter, inflation and the still accelerated cosmic expansion. STQ removes even the initial singularity (infinite energy) of Big Bang theories, but quantum gravity is also required.

Previously, we suggested that the value *a* of the quantum of length could be variable, to reflect the modifiable metric of space and time in general relativity [4]. However, it is simpler and better to assume that *a* is always a universal constant, its value being uniquely determined by the initial energy content of our Universe (hc/2a). Measurements of space-time coordinates could be performed, indeed, by juxtaposing the same quantum of length *a* along any curved geodesic line, followed by photons in gravitational fields. The identity of elementary particles would then be preserved in strong gravitational fields, by defining particle states in terms of phase differences at the smallest possible scale.

## Conclusions

The usual postulate that there exists a "space-time continuum" is not a logical necessity any more, since it is possible to develop a consistent theory, where the smallest measurable distance *a* is a finite universal constant. The continuum assumption was actually a consequence of the classical concept of motions of point-particles along well defined trajectories. Their *continual existence* would prevent that such a particle can move from A to B, without passing through a continuous array of intermediate points. The quantum-mechanical concept of changing probability distributions modified this paradigm in such a way that it can be applied to space-time lattices. Moreover, the theory of relativity prepared the idea that all reference frames have to be equivalent, even for STQ.

Actually, we used these theories as a guide to go one step further. They told us, in particular that physical laws are statements concerning *the knowledge we can get about reality* by means of physically possible measurements. They depend on the existence of universal constants, having specific values. We simply considered three universal constants (c, h and a), where the value *a* of the quantum of length could be finite, instead of imposing that a = 0. This revealed that physics changes as soon as a ≠ 0. Although its actual value is not yet known, it is finite when the total convertible energy content of our Universe is finite, which is at least very probable for Big Bang theories. We also showed that a quantum of length a ≠ 0 would be useful to solve some paradoxes and to get a better understanding of the highly astonishing properties of elementary particles.

In this regard, it was essential to realize that ideally exact measurements of (x, y, z, ct) coordinates in any particular inertial reference frame yield *integer or half-integer multiples of a.* This allows us to distinguish different space-time lattices of lattice constant *a*. The normal lattice contains the origin of the chosen frame, but there are also inserted lattices, displaced by a/2 along one or several reference axes. *States of motion* are defined by means of possible variations of wave functions on anyone of these lattices, but we can also consider a Ψ function for all these lattices. The same probability distribution allows then for phase differences on inserted lattices with respect to the normal lattice, which yields four new quantum numbers. They define *particle states,* so that every type of elementary particles corresponds to a specific pattern of "excitations of space and time at the smallest possible scale". Since there are different types of elementary particles, it was necessary to explain the origin of additional degrees of freedom. The usual concepts of space and time had thus to be modified, but it is not necessary to imagine additional dimensions of space and time, so that elementary particles could be vibrating strings or surfaces. It is sufficient to accept a quantization of ordinary space and time, instead of assuming unobservable realities.

The 3 *color states* of quarks result then from the three-dimensional nature of space, while the 3 *generations* of known elementary particles are related to the time variable. STQ accounts also for other basic ingredients of the "standard model" of elementary particle physics, like the distinction between leptonic and baryonic matter or electro-weak and strong interac-tions. It predicts the existence of yet unknown elementary particles, as Mendeleev's periodic table did for chemical elements. We expect in particular the existence of 8 electrically neutral spin 1/2 fermions, interacting with one another by strong interactions. They can be bound to one another, to form entities that are good candidates for *dark matter particles*. Other particles, allowing for larger symmetry groups, are also possible and could be discovered, when the required energy is sufficient.

The validity of STQ and its predictions concerning dark matter particles can probably be tested in the near future, by means of high-energy collisions at particle accelerators like the LHC, as well as cosmic ray experiments, carried out in satellites.

## References

[1]    W. Heisenberg: Die physikalischen Prinzipien der Quantentheorie, p. 48, 1930.

[2]    A. Meessen: Sur le problème de la quantification de l'espace-temps, Ann. Soc. Sc. Bruxelles **81**, 254-271 (1967).

[3]    A. Meessen: Spacetime quantization, general relativistic mechanics, and Mach's principle. Foundations of Physics **8**, 399-415 (1978).

[4]    A. Meessen: Spacetime quantization, elementary particles and cosmology, Foundations of Physics **29**, 281-316 (1999), http://www.meessen.net/AMeessen/STQ/STQ.pdf

[5]    http://inspirebeta.net/search?ln=en&p=f+t+minimal+length&action_search=Search

[6]    A. Einstein: Autobiographical notes, in Albert Einstein Philosoper-Scientist, P.A. Schilpp, ed. p. 63, 1949.

[7]    A. Einstein, B. Podolsky and N. Rosen: Can quantum-mechanical description of physical reality be considered complete? Phys. Rev. **47**, 777-780 (1935).

[8]    D. Salart, A. Baas, C. Branciard, N. Gizin and H. Zbinden: Testing the speed of 'spooky actions at a distance', Nature **454**, 861-864 (2008).

[9]    Connaissance Scientifique et Philosophie, Roy. Acad. Brussels. 16-17 May, 1973.

[10]    W. Heisenberg: Private communication (1973).

[11]    O. Adriani et al.: Observation of an anomalous positron abundance in the cosmic radiation, arXiv:0810.4995v1 [astro-ph] (2008). An anomalous positron abundance in cosmic rays with energies 1.5-100 GeV, Nature (Letter) **458**, 607-609 (2009).

[12]    T. Aaltonen et al.: Study of multi-muon events produced in $p\bar{p}$ collisions at $\sqrt{s}$ = 1.96 TeV, arXiv:0810.5357v2 [hep-ex]. See also arXiv:1104.0699v1 [hep-ex]

[13]    N. Bornhauser: Analysis of multi-muon signal at collider and fixed-target experiments (2010), http://thp.uni-bonn.de/groups/drees/publ/nicki_susy10.pdf